Routledge
Taylor & Francis Group

ARTICLE

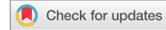

# Algorithmic bias and the New Chicago School

Jyh-An Lee

Faculty of Law, The Chinese University of Hong Kong, Hong Kong SAR, People's Republic of China

**ABSTRACT**
AI systems are increasingly deployed in both public and private sectors to independently make complicated decisions with far-reaching impact on individuals and the society. However, many AI algorithms are biased in the collection or processing of data, resulting in prejudiced decisions based on demographic features. Algorithmic biases occur because of the training data fed into the AI system or the design of algorithmic models. While most legal scholars propose a direct-regulation approach associated with right of explanation or transparency obligation, this article provides a different picture regarding how indirect regulation can be used to regulate algorithmic bias based on the New Chicago School framework developed by Lawrence Lessig. This article concludes that an effective regulatory approach toward algorithmic bias will be the right mixture of direct and indirect regulations through architecture, norms, market, and the law.

**KEYWORDS** Algorithmic bias; automated decision making; artificial intelligence; explainable AI; New Chicago School

## 1. Introduction

Advances in artificial intelligence (AI) algorithms have led to unprecedented progress in data analytics and pattern recognition. Today AI systems are increasingly deployed in both public and private sectors to independently make complicated decisions with far-reaching impact on individuals and the society. While these AI algorithms have become an unprecedentedly efficient and effective tool for identification and assessment, many of them are biased in the collection or processing of data, resulting in prejudiced decisions based on demographic features, such as race, ethnicity, gender, religion, and so on. Some scholars claim that the algorithmic biases in various AI applications have become one of the most urgent and important issues faced by contemporary society.[1] Even tech giants taking the lead in global AI development cannot avoid problems of algorithmic bias created by their own AI systems. For example, in November 2017, Google Translate produced sexist translation from Turkish into English.[2] The Turkish 'O bir doctor' was translated as '*he* is a doctor', whereas 'O bir hemşire' led to '*she* is a nurse'.[3] However, the single-third person pronoun 'O' in Turkish does not mark for gender.[4] This was because the AI in Google Translate learned from the pattern, including gender and occupations, from billions of bilingual texts on the Internet.[5] Moreover, Amazon discontinued its AI-based recruitment system in late 2018 because the system was found biased against women by giving low ratings to resumes with terms, such as 'women's' in the applications for technical positions.[6] Based on the pattern of a decade's worth of resumes from people applying for jobs at Amazon, the company's recruitment AI algorithm taught itself to favour male, which was the gender of most applicants.[7] These are just a few of many examples of how AI embodies biases that have significantly affected our society.

---

[1] See, e.g. Harry Surden, 'Artificial Intelligence and Law: An Overview' (2019) 35 *Georgia State University* Law Review 1305, 1335.

While bias is not a new problem in human society, the deployment of AI in crucial decision making magnifies pre-existing bias or even creates new types of it. In recent years, AI algorithms have been widely deployed to determine the allocation of various resources, such as finance, recruitment, and admissions to educational institutes. When important societal processes such as voting, policing, and sentencing, become increasingly AI-based, these biases are not only technical and social problems, but also legal ones. If technologies in effect displace the values in laws, such as fairness and anti-discrimination, lawmakers need to decide whether and how to reclaim the values displaced.

This article uses Lawrence Lessig's theories of 'code is law' and 'new Chicago School' to understand the ongoing debate concerning AI bias and the dynamic factors reducing or reinforcing algorithmic bias. Part 2 of the article explains how biases are generated by AI algorithms. Algorithms' biases occur because of the inadequate quality and quantity of data, the existing bias in current social structure, or the model design of the algorithms. Part 3 analyses factors that affect algorithmic bias based on the New Chicago School framework of architecture, norms, and market developed by Lessig. Part 4 examines how the law can regulate algorithmic bias through direct regulation and indirectly by regulating architecture, norm, and market. With this analytical framework, this article argues that current literature in law focuses too much on the legal obligation of transparency and explainability. It fails to provide a holistic viewpoint that includes indirect regulation through the other three modalities of regulation. Given the inherit limitation of the legal obligation in transparency and explainability, policymakers should take a multifaceted legislative approach to algorithmic bias. Part 5 concludes.

---

[2]Nikhil Sonnad, 'Google Translate's Gender Bias Pairs 'He' with 'Hardworking' and 'She' with Lazy, and Other Examples' (*Quartz*, 30 November 2017), https://qz.com/1141122/google-translates-gender-bias-pairs-he-with-hardworking-and-she-with-lazy-and-other-examples/?mc_cid=f851a23312&mc_eid=aac87b70fb.
[3]Ibid.
[4]Ibid.
[5]Ibid.
[6]Jeffrey Destin, 'Amazon Scraps Secret AI Recruiting Tool That Showed Bias Against Women' (*Reuters*, 10 October 2018), www.reuters.com/article/uk-amazon-com-jobs-automation-insight/amazon-scraps-secret-ai-recruiting-tool-that-showed-bias-against-women-idUKKCN1MK08K?edition-redirect=uk.
[7]Ibid.

## 2. Bias generated by AI algorithms

Algorithmic biases might originate from the quality of data used to train the AI system, the type of machine-learning method chosen, and the design of the algorithms.[8] A selection bias occurs when the input data does not accurately represent the population or context being analysed.[9] From a statistical viewpoint, the training dataset fails to represent a random sample from the population.[10] This problem exists especially when certain datasets are easier to access for the AI systems but they are not produced based on rigorous statistical standards.[11] This kind of bias is well illustrated in Amazon's recruitment AI algorithm mentioned above, which failed to incorporate sufficient female applicants' data. It has been reported that Apple credit card approval process and the decision of its credit line exhibited similar biased outcomes against female applicants.[12]

Bias also occurs when AI algorithms learn from the data that reflect existing structural inequalities in the society.[13] The outcome from these algorithms will then reinforce undesirable biases and stereotypes.[14] Even if the AI algorithm is not designed to consider sensitive characteristics, such as gender and race, some other correlated features may be used as proxies, which lead to similar biases.[15] Although AI can possibly identify correlations between different variables, it cannot recognise causal connections. For example, because neighbourhoods in the United States (U.S.) are sometimes correlated with race, using 'neighborhood of residence' as a proxy when developing an algorithm for a loan decision may lead to racial biases and controversies.

---

[8] See, e.g. Boris Babic et al., 'When Machine Learning Goes Off the Rails' (*Harvard Business Review*, January 2021), https://hbr.org/2021/01/when-machine-learning-goes-off-the-rails; Miriam C Buiten, 'Towards Intelligent Regulation of Artificial Intelligence' (2019) 10 *European Journal of Risk Regulation* 41, 51–53; Stefan Feuerriegel et al., 'Fair AI: Challenges and Opportunities' (2020) 62(4) *Business & Information Systems Engineering* 379, 381.

[9] See, e.g. Akter et al., 'Algorithmic Bias in Data-Driven Innovation in the Age of AI' (2021) 60 *International Journal of Information Management* 1, 1; Buiten (n 8) 51; Bernhard Waltl, 'Explainability and Transparency of Machine Learning in ADM Systems' in Roland Vogl (ed), *Research Handbook on Big Data Law* (Edward Elgar, 2021) 350.

[10] See, e.g. Akter et al. (n 9) 5.

[11] Eirini Ntoutsi et al., 'Bias in Data-Driven Artificial Intelligence Systems—An Introductory Survey' (2020) 10(3) *WIREs Data Mining & Knowledge Discovery* 1, at 4.

[12] Subrat Patnaik, 'Apple Co-founder Says Apple Card Algorithm Gave Wife Lower Credit Limit' (*Reuters*, 11 November 2019), www.reuters.com/article/us-goldman-sachs-apple-idUSKBN1XL038.

[13] See, e.g. Buiten (n 8) 51–52; Frank Pasquale, *New Laws of Robotics: Defining Human Expertise in the Age of AI* (Harvard University Press 2020) 119; Akter et al. (n 9) 1; Ntoutsi et al. (n 11), 3.

[14] See, e.g. Michael Li, 'Addressing the Biases Plaguing Algorithms' (*Harvard Business Review* 13 May 2019), https://hbr.org/2019/05/addressing-the-biases-plaguing-algorithms.

[15] See, e.g. Buiten (n 8) 53; Waltl (n 9) 342.

Another common origin for bias is the problematic design of the algorithm, which usually leads to over-generalisation or correlation fallacy.[16] Although AI can identify some patterns from the underlying data, it does not understand social context. A notable example is the commercial prediction algorithm, Optum, which helps U.S. hospitals to identify and help patients with significant health needs. Hospitals and health insurers use Optum to identify high-risk patients by looking at their medical histories and their medical expenses, and to detect who need more resources to manage their health. However, because less money was spent on black patients who have the same level of need as white patients, Optum incorrectly determined that black patients were healthier than equally sick white patients.[17] This bias occurred because Optum did not understand that black patients were more concerned about the rising insurance fee resulting from the medical treatment they received, and therefore health costs from insurance data were not a proper measure for health needs.[18]

## 3. The New Chicago School approach to algorithmic bias

Lawrence Lessig advocates for the New Chicago School approach by arguing that behaviour is regulated by four types of constraint: architecture, social norms, market, and the law.[19] These four modalities of regulation, or multiplicity of constraint, operate together to shape human behaviour.[20] This part analyses how AI bias is regulated by these modalities of regulation other than the law.

### 3.1. Architecture

Lessig defines architecture as 'nature' of things, referring to the features of the world made or found by human beings.[21] Architectures regulate behaviours by restricting or enabling them via their nature. In the digital environment, code, including the pervasive software and hardware, is the architecture embedding certain values that regulates behaviour.[22] This is the 'code is law' theory. Algorithmic biases occur because of the selection of data or the design of the algorithms. While these biases lead to discrimination in important social activities, such as hiring, healthcare, insurance, or even criminal justice, the data and algorithms generating these biases present noticeable regulatory power. In this sense, 'code is law' theory provides a proper lens to understand the legal implications of algorithmic biases.

---

[16]See, e.g. Akter et al. (n 9) 8–9.
[17]Ziad Obermeyer et al., 'Dissecting Racial Bias in an Algorithm Used to Manage the Health of Populations' (2019) 366 *Science* 447, 449–451.
[18]Ibid, 452–453.
[19]Lawrence Lessig, 'The New Chicago School' (1998) 27 *Journal of Legal Studies* 661, 662–665.
[20]Ibid, 663.
[21]Lessig (n 19) 666.

Lessig has cautioned that regulation by code is not as transparent as regulation by law.[23] Users of technology are normally not aware of the fact that their behaviours are constrained by the code.[24] They 'experience these controls as nature'.[25] The danger of such code regulation is that it embodies values that significantly affect our life but is not subject to any public scrutiny. This transparency problem of code regulation becomes more serious in AI algorithms. AI applications sometimes covertly amplify social biases with the pretence of objectivity.[26] However, humans can hardly understand AI's decision-making process and predict its outcome.[27] This 'black box' in machine-learning algorithms is inscrutable even for domain experts, especially algorithms incorporating deep-learning or neural-network approaches.[28] Consequently, this opacity obfuscates the discovery of how biases are generated during the operation of AI algorithms.[29] When AI systems are making critical decisions in society, the incompatibility between its black-box nature and the value of due process in modern democracies becomes salient.[30]

To address the 'black box' problem in machine learning, a very common proposal is to push toward more transparency or to implement the principle of explainable AI (XAI) in algorithmic design, a suite of machine-learning techniques that produces more explainable models for humans to understand.[31] XAI allows users to check whether the algorithmic decisions are fair and unbiased[32] by creating the visibility of how an AI system makes decisions and predictions and the possibility to detect flaws in training data and architecture of the algorithms.[33]

---

[22]Lawrence Lessig, 'The Law of the Horse: What Cyberlaw Might Teach' (1999) 113 *Harvard Law Review* 501, 509.
[23]Lawrence Lessig, *Code Version 2.0* (Basic Books, 2006) 138.
[24]See, e.g. Jyh-An Lee and Ching-Yi Liu, 'Forbidden City Enclosed by the Great Firewall: The Law and Power of Internet Filtering in China' (2012) 13 *Minnesota Journal of Law, Science & Technology* 125, 138.
[25]Lessig (n 23) 138; Lessig (n 22) 535, 541.
[26]Karl Manheim and Lyric Kaplan, 'Artificial Intelligence: Risks to Privacy and Democracy' (2019) 21 *Yale Journal of Law & Technology* 106, 111.
[27]Yavar Bathaee, 'The Artificial Intelligence Black Box and the Failure of Intent and Causation' (2018) 31 *Harvard Journal of Law & Technology* 889, 905; Manheim and Kaplan (n 26) 111; Meg Leta Jones and Elizabeth Edenberg, 'Troubleshooting AI and Consent' in Markus D. Dubber et al. (eds) *The Oxford Handbook of Ethics of AI* (Oxford University Press, 2020) 367.
[28]Arun Rai, 'Explainable AI: From Black Box to Glass Box' (2020) 48 *Journal of the Academy of Marketing Science* 137, 137; Waltl (n 9) 344.
[29]Rai (n 28) 137.
[30]Manheim and Kaplan (n 26) 157.
[31]See, e.g. Amina Adadi and Mohammed Berrada, 'Peeking Inside the Black-Box: A Survey on Explainable Artificial Intelligence (XAI)' (2018) 6 *IEEE Access* 52138; Alejandro Barredo Arrieta et al., 'Explainable Artificial Intelligence (XAI): Concepts, Taxonomies, Opportunities and Challenges Toward Responsible AI' (2020) 58 *Information Fusion* 82, 83.

However, the principle of XAI cannot be easily adopted in all types of AI algorithms. With their nested non-linear structure, deep-learning or neural network algorithms normally need to sacrifice transparency and interpretability for prediction accuracy.[34] That is because neural network algorithms are ordinarily not programmed with specific rules that define predictions from the input.[35] These algorithms can automatically find relevant criteria and their decision structures therefore cannot be evaluated.[36] While such machine learning can relieve programmers from writing complicated instructions for computers, its identification of subtle relationships in data might be beyond human's understanding.[37] Explainabilty would then lead to inaccuracy in these deep learning algorithms.[38] Some scholars even argue that a black box model with supra-human accuracy should be compulsory in certain medical fields,[39] rather than being forced to become explainable. Some studies have shown that while it is not technically or economically possible to require ex ante interpretability from deep neural networks, it is often possible for their developers to identify ex post input variables that lead to specific outcomes.[40] However, from a technical viewpoint, the trade-off between explainability and accuracy is always a difficult choice.[41] Therefore, while XAI principle may help decrease algorithmic biases, it is still technically challenging to balance the transparency and accuracy of certain AI algorithms.

Moreover, as mentioned above, a large portion of AI biases result from incomprehensive datasets fed to the learning algorithms. Therefore, data is a major part of the architecture to ensure fair decision making in the AI systems. The fundamental way to solve the problem arising from data is to produce a 'balanced dataset' so that discrimination in the algorithmic outcome can be avoided.[42] AI developers should avoid features such as race, gender or religious affiliation, and other attributes that are highly correlated with these sensitive features.

---

[32]Philipp Hacker et al., 'Explainable AI under Contract and Tort Law: Legal Incentives and Technical Challenges' (2020) 28 *Artificial Intelligence and Law* 415, 429.
[33]Ibid, 429.
[34]See, e.g. Babic et al. (n 8); Heike Felzmann et al., 'Transparency You Can Trust: Transparency Requirements for Artificial Intelligence Between Legal Norms and Contextual Concerns' (2019) *Big Data & Society*, https://doi.org/10.1177/2053951719860542; Rai (n 28) 138.
[35]See, e.g. Buiten (n 8) 50.
[36]See, e.g. Waltl (n 9) 344.
[37]Andrew D. Selbst and Solon Barocas, 'The Intuitive Appeal of Explainable Machines' (2018) 87 *Fordham Law Review* 1085, 1094.
[38]See, e.g. Buiten (n 8) 58; Hacker et al. (n 32) 422.
[39]Hacker et al. (n 32) 423.
[40]Ibid, 417.
[41]Selbst and Barocas (n 37) 1110–1111.
[42]Ntoutsi et al. (n 11) 3.

### 3.2. Norms

There has been rapid development in norms concerning AI ethics. To address the problem of AI bias, there has been a movement in both professional communities and the business sector promoting fairness, accountability, explainability, and transparency in machine learning.[43] The Institute of Electrical and Electronics Engineers (IEEE) published a study, *Ethically Aligned Design,* in which a number of principles, such as human rights, transparency, and accountability, were listed to address the problem of bias generated by AI algorithms.[44] High tech giants, such as Google,[45] IBM,[46] and Microsoft,[47] also released similar AI ethical principles. Several government agencies have published ethics guidelines for more neutral applications of AI as well. For example, the European Commission (EC) published *Ethics Guidelines for Trustworthy AI* in 2019.[48] In October 2020, the European Parliament endorsed the Framework of Ethical Aspects of Artificial Intelligence, Robotics and Related Technologies.[49] The Monetary Authority of Singapore released ethical principles concerning AI and data analytics in the financial sector.[50] The Canadian government also announced major principles in deploying AI in the public sector.[51]

While the above-mentioned AI principles have drawn wide attention from both academia and industries, they have been criticised for being too imprecise, vague abstract, and non-binding.[52] Some research even suggests that the effectiveness of these AI ethical codes is almost zero.[53] Therefore, commentators regard these principles 'toothless'[54] and IEEE has called for clearer legal standards to implement these principles, especially those concerning accountability of algorithmic bias.[55] Compared to the law, norms suffer the disadvantage that they are less clear in terms of boundary and liability.[56] Moreover, an obvious limitation for business norms promoting AI ethics is that companies will not promote norms that benefit the society but put them at a disadvantage compared to their competitors.[57] It is not surprising that some companies use these AI principles for public relations purposes[58] whereas others may even manipulate these principles for their own profit concerns.[59] After all, the norm supreme above all others in the business sector is to maximise shareholder's value or profit.[60] Where companies strive to deploy AI systems, their reasons are also profit-based.[61] Therefore, norms concerning AI ethical principles are not entirely effective in eliminating undesirable algorithmic biases.

---

[43]See, e.g. Pasquale (n 13) 120.
[44]IEEE, *Ethically Aligned Design* (2019) 18–35, https://standards.ieee.org/content/dam/ieee-standards/standards/web/documents/other/ead1e.pdf?utm_medium=undefined&utm_source=undefined&utm_campaign=undefined&utm_content=undefined&utm_term=undefined.
[45]Google, *Artificial Intelligence at Google*, https://ai.google/principles/.
[46]IBM, *AI Ethics*, www.ibm.com/cloud/learn/ai-ethics.
[47]Microsoft, Responsible AI, www.microsoft.com/en-us/ai/responsible-ai?activetab=pivot1:primaryr6.
[48]European Commission, Directorate-General for Communications Networks, Content and Technology, *Ethics guidelines for trustworthy AI*, Publications Office, 2019, https://data.europa.eu/doi/10.2759/177365.
[49]European Parliament, Framework of Ethical Aspects of Artificial Intelligence, Robotics and Related Technologies (20 October 2020), www.europarl.europa.eu/doceo/document/TA-9-2020-0275_EN.html.
[50]Monetary Authority of Singapore, 'Principles to Promote FEAT in the Use of AI and Data Analytics in Singapore's Financial Sector' (2018), www.mas.gov.sg/~/media/MAS/News%20and%20Publications/Monographs%20and%20Information%20Papers/FEAT%20Principles%20Final.pdf.
[51]Government of Canada, 'Responsible Use of Artificial Intelligence (AI)' (12 October 2021), www.canada.ca/en/government/system/digital-government/digital-government-innovations/responsible-use-ai.html.
[52]See, e.g. Inga Strümke et al., 'The Social Dilemma in Artificial Intelligence Development and Why We Have to Solve It', (2021) *AI & Ethics*, https://doi.org/10.1007/s43681-021-00120-w.
[53]Thilo Hagendorff, 'The Ethics of AI Ethics: An Evaluation of Guidelines' (2020) 30 *Minds & Machines* 99, 108.

### 3.3. Market

The market is the main driving force for businesses to develop and use AI. It also affects how businesses approach algorithmic biases. There are different market forces shaping business activities. Firms have incentives to strategically manipulate the algorithmic decision-making process if doing so entails more profit generation.[62] Some social media platforms use algorithmic biases to lock in their users or even manipulate their psychology and behaviour. In this sense, the market is encouraging the production of algorithmic bias.

Nevertheless, market forces driving for more ethical use of AI also develop rapidly. According to Capgemini Research Institute's 2019 survey of 4,400 consumers across six countries, sixty-two percentage indicated that they would place higher trust in a company whose AI interactions they perceived as ethical.[63] Around sixty percentage of the consumers suggested that they would be more loyal to the company and share their positive experience with friends and family.[64] Based on the finding of this research, companies do have an incentive to develop more ethical AI and avoid possible biases in their algorithms so that they can gain trust from consumers. Furthermore, more start-up companies have developed new business models to provide services to certify AI products and processes free from bias, prejudices, and unfairness.[65] Google also offers AI ethics services assisting its clients to examine multiple dimensions of their AI applications, ranging from the data used to train systems to how these systems work to their impact on society.[66] The demand for such ethical or relevant certification services reveals that market competition can also be a positive force alleviating algorithmic biases.

---

[54]Anaïs Rességuier and Rowena Rodrigues, 'AI Ethics Should Not Remain Toothless! A Call to Bring Back the Teeth of Ethics' (2020) 7(2) *Big Data & Society*, https://doi.org/10.1177/2053951720942541.
[55]IEEE (n 44) 19, 29, 40, 103.
[56]Jacqueline D Lipton, 'What Blogging Might Teach About Cybernorms' (2010) 4 *Akron Intellectual Property Journal* 1, 6.
[57]See, e.g. Strümke et al. (n 52).
[58]Hagendorff (n 53) 108.
[59]See, e.g. Rességuier and Rodrigues (n 54).
[60]See, e.g. Mark J. Roe, 'Political Preconditions to Separating Ownership from Corporate Control' (2000) 53 *Stanford Law Review* 539, 554–555.
[61]Hagendorff (n 53) 108.
[62]Selbst and Barocas (n 37) 1093.
[63]Capgemini Research Institute, 'Organizations Must Proactively Address Ethics in AI to Gain the Public's Trust and Loyalty', 5 July 2019, www.capgemini.com/wp-content/uploads/2019/07/2019_07_05_Press-Release_Ethics-in-AI-_EN.pdf.

## 4. Law

While many have argued that legislation is necessary to reduce algorithmic biases,[67] it has been controversial regarding what is the best legal approach to this issue. Lessig distinguishes the New Chicago School from the old school by arguing that law can not only regulate behaviour directly but also regulate it indirectly by regulating market, architecture, and norms.[68] He contends that modern regulation is a mix of these direct and indirect aspects.[69] This argument provides a novel angle to analyse existing laws and legislative proposals concerning algorithmic bias.

### 4.1. Direct regulation

The most straightforward way for lawmakers to address the algorithmic bias is to regulate the behaviour of an AI developer or data processor directly. The law may forbid automated decision making and require human intervention. In *State v. Loomis,* the Wisconsin Supreme Court ruled that when deciding on a sentence, a human judge could consider the scores produced by COMPAS, a risk-assessment algorithm developed by a private company.[70] That was because the sentence was still decided by the judge, instead of the AI algorithm.[71] This human intervention rule was later adopted in Article 22(1) of the European General Data Protection Regulation (GDPR), which provides the data subject with right to object to 'a decision based solely on automated processing, including profiling, which produces legal effects concerning him or her or similarly significantly affects him or her'.[72] The objection right in GDPR has been transplanted in the recent Chinese Personal Information Protect Law (PIPL), which similarly provides data subjects with the right 'to reject the decision made by the personal information processor only through automatic decision-making'.[73] The objection right in Article 22(1) of GDPR is waivable with the data subject's explicit consent.[74]

---

[64]Ibid.
[65]Babic et al. (n 8).
[66]Ibid.
[67]See, e.g. Strümke et al. (n 52).
[68]Lessig (n 19) 666.
[69]Ibid, 666–667.
[70]881 N.W.2d 749 (Wis. 2016).
[71]Ibid, 753.

A more common legislative approach is to mandate certain information disclosure or transparency concerning the algorithmic process. In the above *State v. Loomis* case, the court rejected Loomis's challenge of using COMPAS in sentencing and explained that according to a *Wired* report, 'knowledge of the algorithm's output was a sufficient level of transparency'.[75] The idea to require transparency in algorithms was stipulated in the GDPR in a more detailed way. Article 13 of the GDPR obliges the data controller to provide the data subject with the 'information necessary to ensure fair and transparent processing', which includes 'meaningful information about the logic involved, as well as the significance and the envisaged consequences of such processing for the data subject' in the case of 'automated decision-making, including profiling'.[76]

When a data controller is exempted from Article 22(1) of GDPR, Article 22(3) requires that the controller 'shall implement suitable measures to safeguard the data subject's rights and freedoms and legitimate interests'. In discussing Article 22(3), Recital 71 of the GDPR states that the safeguards should, inter alia, include 'the right … to obtain an explanation of the decision reached after such assessment'. However, since the recital is not binding, most scholars agree that it cannot be a legal basis for a right to an explanation.[77] Article 22 together with Recital 71 has affected Article 24(3) of the recent Chinese PIPL, which likewise provides the data subject with the 'right to request the personal information processor to provide explanation' if he or she believes that 'the automatic decision making has a significant impact on his or her rights and interests'.[78]

Article 15(1)(h) of the GDPR stipulates that in the case of automated processing in the context of Article 22(1), the controller must provide

---

[72]GDPR, art. 22(1).
[73]PIPL, art. 24(3).
[74]GDPR, art. 22(2).
[75]Jason Tashea, "Courts Are Using AI to Sentence Criminals. That Must Stop Now" (*Wired*, 17 April 2017), www.wired.com/2017/04/courts-using-ai-sentence-criminals-must-stop-now.
[76]GDPR, art. 13(2).
[77]See, e.g. Bryce Goodman and Seth Flaxman, 'European Union Regulations on Algorithmic Decision-making and a Right to Explanation' (2017) 38(3) *AI Magazine* 50, 53–54; Andrew D Selbst and Julia Powles, 'Meaningful Information and the Right to Explanation' (2017) 7(4) *International Data Privacy Law* 233, 235; Sandra Wachter et al., 'Why a Right to Explanation of Automated Decision-making Does Not Exist in the General Data Protection Regulation' (2017) 7(2) *International Data Privacy Law* 76, 79–81.
[78]PIPL, art. 24(3).

'meaningful information about the logic involved'. It has been debated by scholars that whether this information only includes the general structure of the model, explanations about individual decisions, or the weight of each variable.[79]

### 4.2. Indirect regulation

Indirect regulation means law constrains behaviour through regulating market, code, or norms. In other words, law uses the regulatory power of other modalities of regulation.[80] The New Chicago School views these other modalities as objects of law's regulation.[81] Based on this understanding, the law can regulate algorithmic bias via market, code, and norms.

#### 4.2.1. Law and market

Law can constrain behaviour through the creation of new market demand. To address the problem of algorithmic bias, some propose a new auditing requirement in the law for AI systems to detect potential biases before these systems are used. This requirement can be found in the Algorithmic Accountability Act, which was introduced to the U.S. House of Representatives in 2019.[82] The idea of such proposals is to reduce algorithmic biases through the creation for a new market for AI auditing.

In fact, some scholars have argued that algorithmic audit is the best way to safeguard against bias generated from the automated decision-making.[83] In recent years, some auditing techniques have been developed to successfully identify algorithmic bias across various industries.[84] Although algorithmic audits may be voluntarily conducted by AI development companies or their clients without the law,[85] the law might help to further strengthen the effectiveness of such auditing.

While the GDPR has authorised government authorities to conduct data auditing to combat algorithmic discrimination in opaque AI systems,[86] a more effective auditing approach might be requiring accounting firms or other private accredited professional organisations to conduct the algorithmic audit. In that way, the law can promote algorithmic fairness through a new market of algorithmic auditing and the norms emphasising independence in the professional auditing community. The law does not necessarily need to make algorithmic auditing compulsory.[87] Instead, it may choose to encourage such auditing by limiting AI companies' liability or providing them with priority in government procurement projects or other public resources. Although it is still challenging to develop an effective auditing approach for algorithmic audits,[88] using the law to encourage the market to develop more comprehensive auditing techniques can certainly help alleviate the problem of algorithmic bias.

---

[79]See, e.g. Gianclaudio Malgieri and Giovanni Comandé, 'Why a Right to Legibility of Automated Decision-making Exists in the General Data Protection Regulation' (2017) 7(6) *International Data Privacy Law* 243, 246–247; Selbst and Powles (n 77) 241; Wachter et al. (n 77) 83.
[80]Lessig (n 19) 666.
[81]Ibid.
[82]H.R.2231 - Algorithmic Accountability Act of 2019, 116th Congress of the United States (2019–2020).
[83]Bryan Casey et al., 'Rethinking Explainable Machines: The GDPR's "Right to Explanation" Debate and the Rise of Algorithmic Audits in Enterprise' (2019) 34 *Berkeley Technology Law Journal* 143, 152, 182.
[84]Casey et al. (n 83) 182.
[85]Wolfgang Hoffmann-Riem, 'Artificial Intelligence as a Challenge for Law and Regulation' in Thomas Wischmeyer and Timo Rademacher (eds) *Regulating Artificial Intelligence* (Springer, 2020) 19.
[86]Chapter 6, GDPR.

### 4.2.2. Law and code

The law may use code, or architecture, to bring about a regulatory change.[89] In other words, instead of regulating behaviour directly, the law can choose to regulate how algorithms are designed. GDPR has taken this 'data protection by design' approach in some of its rules.[90] Its Recital 71, according to which processing under Article 22, requires data controllers to 'use appropriate mathematical or statistical procedures for the profiling … and prevent, inter alia, discriminatory effects on natural persons on the basis of racial or ethnic origin, political opinion, religion or beliefs, trade union membership, genetic or health status or sexual orientation, or processing that results in measures having such an effect'.[91] This provision can be understood as law regulating the code by requiring the later to incorporate 'appropriate mathematical or statistical procedures'.

Another example of law's regulating code concerning AI algorithms is recent EC's proposal of the AI Act, which requires '[h]igh-risk AI systems shall be designed and developed in such a way to ensure that their operation is sufficiently transparent to enable users to interpret the system's output and use it appropriately'.[92] While there might be some controversy regarding to what extent a user can interpret the output from an AI system, this legislative approach aims to regulate behaviour via requiring interpretability in AI algorithmic design.

---

[87]Hoffmann-Riem (n 85) 19.
[88]Adriano Joshiyama et al., 'Towards Algorithm Auditing: A Survey on Managing Legal, Ethical and Technological Risks of AI, ML and Associated Algorithms', 15 February 2021, https://papers.ssrn.com/sol3/papers.cfm?abstract_id=3778998.
[89]Lessig (n 22) 512.
[90]Casey et al. (n 83) 179–183.
[91]GDPR, Recital 71.
[92]European Commission, Proposal for a Regulation of the European Parliament and of the Council, Laying Down Harmonised Rules on Artificial Intelligence (Artificial Intelligence Act) and Amending Certain Union Legislative Acts, 21 April 2021, art. 13(1), https://eur-lex.europa.eu/legal-content/EN/TXT/HTML/?uri=CELEX:52021PC0206&from=EN.

### 4.2.3. Law and norms

The relation between norms and law has been a major focus of the New Chicago School.[93] Although both business sectors and governments have advocated for principles regarding AI ethics, their motives for doing so are quite different. Businesses express their support for ethical AI development because they want to gain trust from their customers. Some even believe that by embracing AI ethical guidelines, AI companies are delivering a message to the policymakers that no legislation is needed in this area.[94]

Policymakers can incorporate non-legal norms to encourage or discourage certain behaviours.[95] As mentioned above, several government agencies have promoted norms based on ethical AI principles. In addition to stimulating the social awareness of algorithmic bias, these government initiatives are a foundation for the society to discuss the necessity and substance of future legislation. A noteworthy example is the European Union (EU)'s approach to AI legislation. Since the High-Level Expert Group set up by the EC first issued the *Ethics Guidelines for Trustworthy AI* in 2019,[96] the EU has published a series of principles associated with AI ethics,[97] which all become the basis of the EC's legislative proposal for a regulation on Artificial Intelligence, the so-called 'Artificial Intelligence Act' or 'AI Act', in April 2021.[98] This development illustrates how norms are codified in the law if they are able to penetrate into the consciousness of a large portion of the population,[99] and how governments promote social norms to build social consensus for future legislation.

Social norms are usually enforced through the social structure of a community where an individual lives,[100] and law can indirectly regulate behaviour through supporting this structure.[101] Given the pervasive deployment of AI and the increasingly critical bias problem, a plausible proposal is to use the law to establish professional societies for AI developers. The law can require all AI developers to join the society, and unethical behaviour could lead to disqualification from it.[102] One advantage of this proposal is that a professional society has more power than individual AI developers in insisting on AI principles against the employer's commercial interests. Furthermore, compared to the government or companies, a professional society can develop and enforce the AI ethics code through the pride of a profession. The law can also mandate or encourage certification of AI ethics compliance developed by professional organisations, such as IEEE or International Organization for Standardization. With the support of the law, the community of professional AI developers can better work collectively toward more ethical AI development.

---

[93] See, e.g. Mark Tushnet, 'Everything Old Is New Again: Early Reflections on the "New Chicago School"' (1998) 1998 *Wisconsin Law Review* 579, 579.
[94] Hagendorff (n 53) 100.
[95] Tushnet (n 93) 581.
[96] European Commission (n 48).
[97] Part 3.2 of this article.
[98] European Commission (n 92).
[99] See, e.g. Cass R. Sunstein, 'Social Norms and Social Roles' (1996) 96 *Columbia Law Review* 903, 915.
[100] Lessig (n 19) 662; Lipton (n 56) 6.
[101] Ibid. at 669.
[102] Strümke et al. (n 52).

Law can also reinforce ethical AI applications through organisational structure and culture. Some commentators argue that companies' internal governance has failed to assure the fairness of AI decision making.[103] To address algorithmic bias and other relevant AI ethical issues, some companies have introduced AI ethics committees as an internal governance framework to examine how they design and implement AI systems.[104] Although these initiatives are profit-oriented with the view to mitigate business risks,[105] this is a positive development for more ethical and fair use of AI. The idea of an AI ethics committee comes from the legal requirement of the committee-based oversight for all research involving humans and vertebrate animals, which might lead to ethical concerns.[106] Currently, there has been no consensus on the composition of the committee, the scope of its review, and the level in the company. However, once more companies establish similiar committees in their organisations, the law might reinforce such ethics monitoring mechanism by making the committee a mandatory part of the corporate organisation.

The legal evolution of independent directors provides a good example for a possible future of AI ethics committees. Independent directors were viewed as a 'good governance' exhortation, which companies could adopt on a voluntary basis in the United States.[107] Companies had started to have more incentives to include independent directors in their boards since the Delaware courts applied the lenient 'business judgment rule' to board action undertaken by independent directors.[108] Gradually, independent directors have become a mandatory element of company and securities law, especially for public-listed companies.[109] The development of the independent director mechanism provides an insightful lesson for AI ethics committee. Policymakers may provide enterprises with some incentives for the voluntary adoption of AI ethics committees, such as the reduction of legal liability for companies involved in litigation associated with AI ethics dispute or priorities in government's AI systems procurement. Once a critical mass of AI companies has instituted such an organisation design, the legislature can consider making it a compulsory governance mechanism under some circumstances.

---

[103]Feuerriegel et al. (n 8) 383.
[104]See, e.g. Zara Stone, 'The Artificial Intelligence Ethics Committee' (*Forbes* 11 June 2018), www.forbes.com/sites/zarastone/2018/06/11/the-artificial-intelligence-ethics-committee/?sh=375a0baf1637;
Steven Tiell, 'Create an Ethics Committee to Keep Your AI Initiative in Check'(*Harvard Business Review* November 2019), https://hbr.org/2019/11/create-an-ethics-committee-to-keep-your-ai-initiative-in-check.
[105]Tiell (n 104).
[106]Accenture and Northeastern University Ethics Institute, *Building Data and AI Ethics Committees* (2019) 7, www.accenture.com/_acnmedia/PDF-107/Accenture-AI-And-Data-Ethics-Committee-Report-11.pdf#zoom=50.
[107]Jeffrey Gordon, 'The Rise of Independent Directors in the United States, 1950-2005: Of Shareholder Value and Stock Market Prices' (2007) 59 *Stanford Law Review* 1465, 1468.
[108]Ibid.
[109]Ibid.

### 4.3. Information disclosure and algorithmic bias

Most current laws and legislative proposals regulate algorithmic bias through different levels of information disclosure requirement. All direct and indirect regulations of algorithmic biases described above aim to provide more transparency or information about the automated decision-making process. Although the legal requirement of transparency and information disclosure will not directly eliminate algorithmic biases, they can certainly enhance the trustworthiness of AI systems and provide stakeholders with more opportunities to evaluate the validity and justifiability of the automated decision making.[110]

However, from a technical perspective, there are different possible explanations of transparency in AI systems.[111] From a legal viewpoint, there has been no consensus regarding what information should be disclosed, to what extent it should be disclosed, and how it should be disclosed required by the law.[112] Take the GDPR for example, scholars are still debating whether there is a right of explanation for data subject and whether the law requires an ex ante explanation of how the system functions, the so-called 'prospective transparency',[113] or an ex post explanation of each individual automated decision,[114] the so-called 'retrospective transparency'.[115] This article does not intend to join the legal debate regarding transparency, explainability, or interpretability. Instead, it aims to identify the key technical and legal challenges for using transparency as a major liability mechanism in regulating algorithmic biases.[116]

Scholars have argued that algorithms without explainability or transparency should be prohibited from making important decisions, such as credit rating.[117] If the policy goal is to avoid algorithmic biases, the transparency or disclosure requirement should allow stakeholders to trace back the factors or variables that lead to certain outcomes in the decisions.[118] In other words, stakeholders should be able to understand how changing a certain factor would affect the final decision. However, a legal requirement for algorithmic transparency is not a panacea to the issue arising from AI's 'black box' approach. While the costs in making training and testing data transparent might be acceptable to AI companies, transparency in AI's decision-making process is sometimes economically and technically infeasible for them. In addition to the technical dilemma between AI's accuracy and explainability, the law needs to consider whether individual users can really understand the legally required explanation. Stakeholders normally fail to understand how an algorithm makes certain decisions even if its source code is made available.[119]

---

[110]Felzmann et al. (n 34).
[111]Nicholas Diakopoulos, 'Transparency' in Markus D. Dubber et al. (eds.), *The Oxford Handbook of Ethics of AI* (Oxford University Press, 2020) 200.
[112]See, e.g. Buiten (n 8) 45.
[113]Felzmann et al. (n 34).
[114]Maja Brkan, 'Do Algorithms Rule the World? Algorithmic Decision-Making and Data Protection in the Framework of the GDPR and Beyond' 27(2) *International Journal of Law and Information Technology* 91, 110–111; Selbst and Barocas (n 37) 1107–1108, 1122–1123.
[115]Felzmann et al. (n 34).
[116]Felzmann et al. (n 34).
[117]Waltl (n 9) 344.

Furthermore, although information disclosure or transparency can help stakeholders to identify possible algorithmic biases, this individual remedy is not a perfect solution. Transparency in algorithms will not necessarily lead to the knowledge about whether and how biases are generated from them.[120] Individual users may lack the capability to understand the complicated process in the AI systems even if they are provided with relevant information.

Moreover, with the logic and model processing details underlying the AI systems at hand, individual users may still fail to see the possible bias without a bird's eye view at the statistical scale.[121] However, legislators need to consider that one of the positive outcomes of legally required explanation is that it may eventually increase digital literacy of the public.[122] In other words, the law may help individual users to understand more about algorithmic decision–making processes and underlying variables as a result of adopting a reasonable scope of explanation or information disclosure obligation. This education function will partly help to resolve the algorithmic bias problem.

Imposing a transparency requirement on AI algorithms may create a tension within the legal system. Source code of computer programmes and other relevant information are trade secrets or confidential information in most jurisdictions. Therefore, disclosure of the algorithmic model and logic means the software proprietor needs to give up its trade secret over algorithms.[123] Policymakers need to consider to what extent that trade secret protection should compromise with policy concerns of transparency in algorithms.

---

[118]See, e.g. Felzmann et al. (n 34).
[119]See, e.g. Buiten (n 8) 54; Selbst and Barocas (n 37) 1094.
[120]Selbst and Barocas (n 37) 1094.
[121]Casey at el. (n 83) 181.
[122]Jones and Edenberg (n 27) 368.
[123]See, e.g. Diakopoulos (n 111) 200; Selbst and Barocas (n 37) 1092–1093.

While transparency in AI algorithms is desirable for stakeholders to understand whether and how biases come from, imposing a wild legal requirement of transparency is certainly not a smart move. When considering such a requirement, legislators should contemplate what kind of transparency is most useful and technically available. If the policy goal is to maintain the fairness of algorithmic decision making, explainability or information disclosure is the means rather than the end.

Given the above-mentioned legal controversies over the definitions of explainability and transparency, policymakers should consider using indirect regulation to foster a more feasible and effective monitoring mechanism for AI algorithms. Through the new auditing market and professional AI developer societies introduced previously, a more practical code for AI governance can be possibly developed through private ordering. The law does not need to intervene in this unclear water with every detail. The successful development and wide application of the generally accepted accounting principles (GAAP) is such an example, where government can rely on this professional guideline developed through the collective action of professional bodies.

## 5. Conclusion

Biases have been part of human society for a long time and we can never eliminate them with the law and other modalities of regulation. With AI systems increasingly making various critical decisions in our society, algorithmic bias has been an important social and legal issue. Most current literature addressing the problem of algorithmic bias focuses on the direct regulation associated with users' right of explanation and AI operators' transparency obligation. Recognising the significance of this direct regulation, this article identifies the limitations of this approach and provides a different picture of regulating AI bias via code, norms, and market. The understanding of the interactions between these modalities will help policymakers ponder what is the best regulatory strategy against the fast-changing AI technologies and the subsequent bias problem. This article concludes that an effective regulatory approach toward algorithmic bias will be the right mixture of these modalities.

## Acknowledgments

The author is grateful to Roger Brownsword, David Donald, and Tianxiang He for their insightful comments and suggestions. This article has also benefited from feedback provided in the Law, Technology, and Disruption Conference held by the City University of Hong Kong School of Law, and Technology and the Evolution of Law Seminar held by the Chinese University of Hong Kong Faculty of Law. The author also thanks Yangzi Li and Jingwen Liu for their research assistance.

## Disclosure statement

No potential conflict of interest was reported by the author(s).

## Notes on contributor

*Jyh-An Lee* is a Professor and Executive Director of the Centre for Financial Regulation and Economic Development (CFRED) at The Chinese University of Hong Kong Faculty of Law.